\documentclass[journal,onecolumn]{IEEEtran}
\usepackage{float}
\usepackage{cite}
\ifCLASSINFOpdf
   \usepackage[pdftex]{graphicx}
   \DeclareGraphicsExtensions{.pdf,.jpeg,.png}
\else
\fi
\usepackage[cmex10]{amsmath}
\begin{document}
\title{Automatic Transcription of Flamenco Singing from Polyphonic Music Recordings}

\author{Nadine~Kroher
        and~Emilia~G\'{o}mez
\thanks{N. Kroher and E. G\'{o}mez are with the Music Technology Group
(MTG) of the Department of Information and Communication
Technologies, Universitat Pompeu Fabra, Barcelona, Spain, e-mail:
\{nadine.kroher,emilia.gomez\}@upf.edu}% <-this % stops a space
}


\markboth{IEEE TRANSACTIONS ON AUDIO, SPEECH, AND LANGUAGE PROCESSING}%
{Kroher \MakeLowercase{\textit{et al.}}: Automatic Transcription of Ornamented Singing from Polyphonic Music Recordings}




\maketitle

\begin{abstract}
Automatic note-level transcription is considered one of the most challenging tasks in music information retrieval. The specific case of flamenco singing transcription poses a particular challenge due to its complex melodic progressions, intonation inaccuracies, the use of a high degree of ornamentation and the presence of guitar accompaniment. In this study, we explore the limitations of existing state of the art transcription systems for the case of flamenco singing and propose a specific solution for this genre: We first extract the predominant melody and apply a novel contour filtering process to eliminate segments of the pitch contour which originate from the guitar accompaniment. We formulate a set of onset detection functions based on volume and pitch characteristics to segment the resulting vocal pitch contour into discrete note events. A quantised pitch label is assigned to each note event by combining global pitch class probabilities with local pitch contour statistics. The proposed system outperforms state of the art singing transcription systems with respect to voicing accuracy, onset detection and overall performance when evaluated on flamenco singing datasets. 
\end{abstract}


% \fi
\IEEEpeerreviewmaketitle

\section{Introduction}
\subsection{Definition and Motivation}
\IEEEPARstart{F}{lamenco} music is a rich improvisational art form with roots in Andalusia, a province in southern Spain. Due to its particular characteristics and importance for the cultural identity of its area of origin, flamenco as an art form was inscribed in the UNESCO List of Intangible Cultural Heritage of Humanity in 2010. Given the growing community of flamenco enthusiasts around the world, a need for computational methods to aid study and diffusion of the genre has been identified. First efforts have been made to adapt existing music information retrieval (MIR) techniques to the specific nature of flamenco \cite{BRIDGES}. Having evolved from an a cappella singing tradition \cite{FLA2}, the vocals represent the central element of flamenco music, accompanied by the guitar, percussive hand-clapping and dance. Consequently, the main focus is set on developing algorithms which target the analysis of the singing voice. 

Flamenco is a strongly improvisational and sparsely documented oral tradition, where songs and techniques have been passed from generation to generation. Given the resulting lack of scores, studies often rely on scant, labour-intensive manual transcriptions. In this study, we present a novel system for automatic singing transcription from polyphonic music recordings targeting the particular case of flamenco. The resulting automatic transcriptions are essential to a number of related MIR tasks, such as melodic similarity characterisation \cite{SIM}, similarity-based style recognition \cite{FITTING}, singer identification \cite{SINGER-ID} or melodic pattern retrieval \cite{PATTERN} and can furthermore aid a broad variety of musicological studies \cite{ETHNO}.

Automatic singing transcription (AST) refers to the extraction of a note-level representation corresponding to the singing melody directly from audio recordings. This task comprises two main challenges: First, the estimation of the fundamental frequency corresponding to the sung melody ({\em vocal pitch extraction}) and second, its conversion into discrete note events ({\em note transcription}). In the resulting symbolic representation, each note is described by its onset time, duration and a pitch value quantised to the equal tempered scale. While the generalised task of automatic music transcription (AMT) \cite{AMTold, AMTnew} is considered a major challenge in MIR, the genre under study poses a number of additional difficulties for both, the vocal pitch extraction and the note transcription stage. Pitch estimation is a well-studied problem in MIR with a variety of sub-tasks such as multi-pitch and predominant melody extraction \cite{PITCH}. The difficulty in the context of flamenco singing is to extract the pitch contour corresponding to the vocal melody while omitting contour segments which originate from the guitar accompaniment. The voice is usually not present throughout the entire song but alternates with interludes in which the guitar takes over the main melodic line. Note transcription can be a trivial task when, depending on the instrumentation, note onsets coincide with significant pitch and volume discontinuities. The singing voice on the other hand, given its non-percussive and pitch-continuous nature, poses a particular challenge when segmenting a pitch contour into discrete note events. Flamenco singing is characterised by a large amount of melodic, partly micro-tonal, ornamentations, excessive use of vibrato and instability of timbre and dynamics. Furthermore, vocal melodies are often composed of a succession of conjunct degrees and singers tend to intonate significantly above or below target notes \cite{FLA1}. Consequently, the complexity of onset detection and pitch labelling significantly increases and requires an algorithmic design which considers the aforementioned characteristics.

\subsection{Related work}
Approaches to automatic note-level transcription from audio recordings date back as far as 1977 \cite{MOORER77} and are extensively reviewed by \cite{AMTold} and \cite{AMTnew}. Most systems described in these reviews provide a generic transcription framework covering a wide range of instruments and musical genres. As mentioned in \cite{AMTnew}, specific instruments exhibit particular characteristics which might not be captured by a generic transcription system. 

Approaches dealing with the specific task of singing transcription have usually been developed in user-oriented MIR tasks such as query-by-humming (QBH), query-by-singing (QBS), sight-singing tutors or computer-aided composition tools. Such systems mainly assume user-input, i.e. the user singing a query, and are consequently designed to transcribe unaccompanied recordings sung by amateur singers. In this case, the vocal pitch extraction stage of the transcription algorithm is reduced to a monophonic pitch estimation problem and systems mainly differ in the note segmentation stage: \cite{ANTONELLI08} and  \cite{POLLASTRI02} obtain an initial detection of note onsets directly from discontinuities in the volume envelope and \cite{WANG03} use a detection function based on changes in spectral band energies. There are obvious limitations to these rather simple segmentation approaches: Consecutive notes sung in legato may have a stable volume envelope and accompanying instruments may cause sudden spectral variation without a singing voice onset being present. A first approach entitled {\em island building} based on pitch contour characteristics was proposed in \cite{MCNAB97} and still finds application in more recent systems \cite{ANTONELLI08, POLLASTRI02}: Consecutive frames with a pitch estimate within a limited range are grouped into notes events. In a comparative study \cite{ADAMS06} a variety of pitch contour based segmentation approaches are compared, including adaptive filters, maximum likelihood estimation, probabilistic modelling and local quantisation. A recent approach to note segmentation \cite{SIPTH}, formulates interval transition as a hysteresis curve and locates note onsets based on the local cumulative pitch deviation. 

Addressing the more complex task of singing transcription from polyphonic recordings \cite{RY06}, a multiple fundamental frequency estimator with an accent signal in order to extract the dominant pitch trajectory. The segmentation stage relies on a probabilistic note event model, using a Hidden Markov Model (HMM) trained on manual transcriptions. In an extension to this approach \cite{KRIGE08}, note transition probabilities were incorporated in the computational model. Recently, this note segmentation method was implemented in the computer-aided note transcription tool {\em tony}\cite{TONY}. It should be mentioned that such probabilistic methods require a large amount of ground truth data during the training stage. This involves time-consuming manual annotations in particular for improvisational and strongly ornamented singing performances, as it is the case for flamenco music. 

A recent trend in music information retrieval has focused on culture-specific non-Western music traditions \cite{SERRA} and has led to the adaptation of existing MIR approaches as well as the development of novel genre-specific techniques. It has been shown that underlying musicological assumptions, i.e. regarding rhythm or tonality, do not necessarily hold for non-Western music styles. In the context of flamenco singing, a first approach for monophonic transcription was proposed in \cite{FITTING}: Based on a contour simplification algorithm \cite{FITTING2}, an estimated monophonic pitch track is converted into a set of constant segments within which the absolute error between the pitch track and the fitted constant does not exceed a pre-determined threshold. A system for computer-assisted transcription of single-voiced a cappella singing recordings has been proposed in \cite{MONO}: Given the absence of accompaniment, a monophonic pitch estimator based on spectral auto-correlation (SAC) represents the front-end of the system. The note segmentation stage is based on a likelihood maximisation method \cite{PATENT}: A dynamic programming (DP) algorithm is used to find the best among all possible notes segmentations along the entire track. The resulting short note transcription is refined in a post-processing stage, containing an iterative tuning estimation and short note consolidation process. The system was extended \cite{POLY} to the transcription of sung melodies from polyphonic flamenco recordings by replacing the monophonic pitch estimator with a predominant pitch extraction algorithm \cite{MELODIA}. 

Nevertheless, the authors report mistakenly transcribed guitar notes as a main source of error and the inaccurate vocal detection a the major limitation of the system performance. The reported note transcription accuracies reported in \cite{MONO} with a note f-measure of slightly below $0.4$ furthermore indicate the difficulty of transcribing flamenco singing. Significantly higher performance is achieved when evaluating the same transcription algorithm on a dataset containing pop- and jazz singing excerpts. We therefore identify a need for improving the state of the art on flamenco singing transcription and design an algorithm robust towards the particular characteristics of the genre which is furthermore suitable for the analysis of accompanied flamenco singing recordings.

\subsection{Contributions and paper outline}
We propose a novel AST system capable of transcribing strongly ornamented improvisational flamenco performances and achieving higher accuracies than state of the art methods. Similar to G{\'o}mez {\it et al.} \cite{POLY}, we use a predominant melody extraction algorithm to extract the vocal pitch contour. We extend this method in two ways: In a pre-processing stage, we use spectral characteristics to select the stereo channel in which the vocals are more dominant for further processing. Subsequent to the predominant melody extraction, we apply a novel contour filtering method, in which pitch contour segments corresponding to guitar melodies are eliminated. In the note transcription stage, we formulate a set of novel onset detection functions based on pitch contour and volume characteristics, which prove robust towards the influence of the accompaniment as well as a high degree of melodic ornamentations. A pitch label is assigned to each note event by combining overall chroma probabilities with local pitch statistics. We refine the resulting symbolic transcription by applying musicological constraints regarding note duration and pitch range. 

The remainder of the paper is organised as follows: We provide a detailed description of the proposed system in Section \ref{sec:method} and specify the evaluation methodology in Section \ref{sec:evaluation}. We give experimental results in Section \ref{sec:experiments} and finally conclude the study in Section \ref{sec:conclusions}.

\section{Proposed Method}
\label{sec:method}

\begin{figure}
\begin{minipage}[b]{1.0\linewidth}
  \centering
  \centerline{\includegraphics[width=6.5cm]{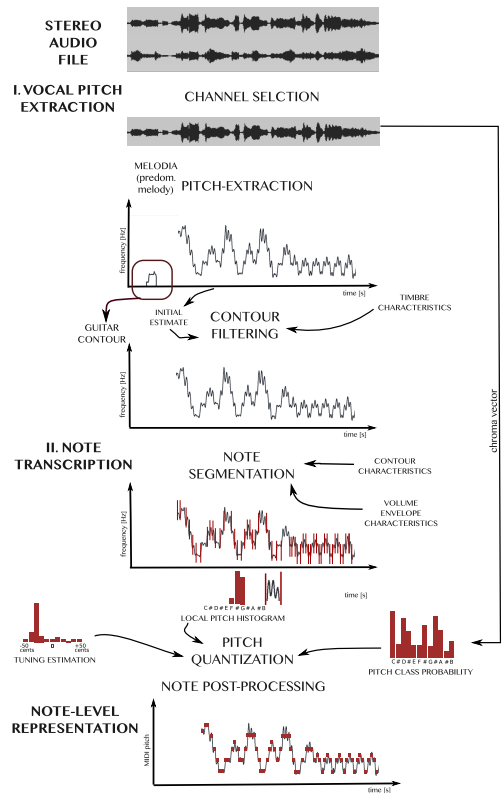}}
\end{minipage}
\vspace*{-0.3cm}
\caption{Block diagram of the proposed system.}
\label{fig:sysOver}
\end{figure}

The processing blocks of the proposed system are depicted in Figure \ref{fig:sysOver}. As mentioned above, automatic singing transcription systems consist consist of two main processing blocks: A vocal pitch extraction algorithm and a note transcription stage. Both stages of the proposed algorithm are described in detail below. 

\subsection{Vocal Pitch Extraction}
\label{sec:vocal melody extraction}
In the context of singing voice transcription, we use the term {\em voicing} to describe whether the vocals are present in a frame or not. Consequently, for both monophonic and polyphonic transcription systems, the pitch values of {\em unvoiced} frames need to be eliminated in order to obtain an accurate time-frequency representation of the vocal pitch. For monophonic singing transcription systems, the task of vocal pitch extraction is reduced to a monophonic pitch estimation problem, since the singing voice represents the only instrument present. Nevertheless, silences and background sounds can cause noisy sections in the pitch track during unvoiced frames. While \cite{ANTONELLI08} and \cite{POLLASTRI02} locate unvoiced frames solely based on the volume envelope, in \cite{SIPTH} a tree classifier based on a set of low-level descriptors in used. For polyphonic recordings, where the singing voice is accompanied by one or more instruments, the task of vocal pitch extraction gains in complexity due to the presence of various harmonic sources. In flamenco music, the singing voice is the dominant element accompanied by the guitar, which takes over the main melodic line during interludes. An obvious approach to extracting the vocal pitch would be to isolate the voice signal by applying source separation methods and using a monophonic pitch estimator on the resulting voice signal. A different strategy was applied in \cite{POLY}, where a predominant melody extraction algorithm is used to estimate the vocal melody. This method gives convincing results for sections of a track where the vocals represent the dominant music element. The same study has furthermore shown that the obtained overall performance regarding voicing and raw pitch accuracy is superior to a source separation approach. Nevertheless, mistakenly transcribed guitar contour segments during interludes, where the voice is absent and the guitar takes over the main melodic line, cause a relatively high number of voicing false positives. 

Based on these prior findings, we adopt the predominant melody approach and extend it in order to reduce the number of mistakenly transcribed guitar pitch contour segments. We first exploit the fact that in flamenco stereo recordings, the vocals are usually more dominant in one of the two channels and use only this channel for further processing. After extracting the predominant melody, we apply a novel frame-based vocal / non-vocal classification and eliminate parts of the pitch contour which are located outside the vocal sections. Vocal detection has been explored mainly as a machine learning tasks \cite{VD1,VD2,VD3}. While good results have been obtained on a frame-level, such algorithms require a large amount of manually annotated training data. Here, we propose a novel vocal detection algorithms which works on a track level and eliminates guitar contour segments based on spectral characteristics.

\subsubsection{Channel selection}
\label{ssec:channel selection}
In pop music productions, it is common practice to place the vocals in the centre of the panorama during the mixing process. In flamenco recordings on the other hand, the vocals often appear stronger in one stereo channel. One reason for this phenomenon is that many such recordings are live stereo recordings, where the resulting panorama distribution corresponds to the physical location of singer and guitarist on stage. Even in multi-track flamenco productions it is a tradition to separate the voice and guitar in the artificially created panorama. An exception are productions with an extended instrumentation, i.e. two guitars or additional instruments, which is rather uncommon, or the obvious case of mono recordings. We want to exploit this fact by automatically selecting the channel with the stronger presence of the vocals for further processing in order to reduce the influence of the guitar during the vocal pitch extraction stage. We would like to point out, that while this channel selection improves system performance as shown in Section \ref{sec:experiments}, our system does not rely on having a panorama separation of the sources. We simply exploit this fact whenever it is the case for a given track. 

\begin{figure}
\begin{minipage}[b]{1.0\linewidth}
  \centering
  \centerline{\includegraphics[width=7.3cm]{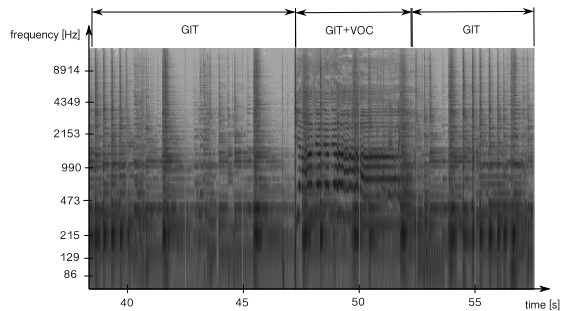}}
\end{minipage}
\vspace*{-0.3cm}
\caption{Spectrogram for voiced and unvoiced sections: "GIT" denotes the presence of the guitar; "VOC" denotes the presence of the singing voice.}
\label{fig:channelSelection}
\end{figure}

The automatic channel selection is based on the fact that guitar and voice differ in their distribution of spectral energy (Figure \ref{fig:channelSelection}): When the vocals are present, we observe an increase the frequency range of 500Hz to 6kHz. We therefore select the stereo channel with a higher average presence in this range. For both audio channels with a sampling rate of $f_s=44.1$ kHz, we compute the Short Time Fourier Transform (STFT) for a window size $N=4096$ samples, a zero padding factor of $m=2$ and a hop size of $h_s=1024$ samples. Accordingly, the bin $k$ corresponding to a centre frequency $f$ is given as

\begin{equation}
\label{bins}
k(f)=\mathrm{round} \left(\frac{f\cdot m \cdot N}{f_s}  \right)
\end{equation}
To characterise the presence of the voice, we compute the spectral band ratio $S[n]$ from the ratio of summed magnitudes in the upper band $f_{21}=500\mathrm{Hz} < f < f_{22}=6000\mathrm{Hz}$ to the lower band $f_{11}=80\mathrm{Hz} < f < f_{12}=400\mathrm{Hz}$:
\begin{equation}
\label{SBR}
S[n]=20 \cdot \log 10\left( \frac{\sum\limits_{k(f_{21})<k<k( f_{22})}|\dot{X}[k,n]|}{ \sum\limits_{k(f_{11})<k<k(f_{12})}|\dot{X}[k,n]|} \right)
\end{equation}
In order to eliminate the influence of the overall volume, we divide the frame magnitude spectrum $|{X}[k,n]|$ by its maximum value, resulting in the normalised magnitude spectrum $|\dot{X}[k,n]|$ at time frame $n$. We compute the frame-wise spectral band ratio $S[n]$ and average for each channel separately over the entire length of the song. Based on the resulting two average spectral band ratios, $\overline{S_{\mathrm{left}}}$ and $\overline{S_{\mathrm{right}}}$, we select the channel with the higher average value. Preliminary experiments have shown that comparing the summed bin magnitude values $|X[k,n]|$ yields to a better discrimination than comparing the summed signal energy $|X[k,n]|^2$ among the two bands. 

\subsubsection{Predominant melody extraction}
\label{ssec:Predominant melody extraction}

Predominant melody extraction refers to the task of estimating the pitch track corresponding to the perceptually dominant melody in a given polyphonic music recording. It furthermore includes the task of determining if the main melody is present in a given time frame or not. In the scope of this paper we will refer to frames in which the main melody is estimated to be present as {\em melody frames} and all remaining frames as {\em non-melody frames}. Furthermore, we define a {\em contour} as a sequence of consecutive melody frames. We extract the predominant melody from the previously selected channel using the algorithm proposed in \cite{MELODIA}. We selected this particular method due its available implementation in the {\em essentia} library \cite{ESSENTIA} and in order to obtain a direct comparison to the study presented in \cite{POLY}, where it was also used as the front end of a flamenco singing transcription. We would like to mention that this method can be replaced by any other predominant melody extraction algorithm. While in the general task of pitch extraction, the performance is mainly limited by pitch estimation errors, i.e. octave errors, the limitation in the scope of vocal pitch estimation is to a large extend conceptual: Even a perfect predominant melody extraction algorithm will estimate contours originating from the guitar, since the guitar accompaniment represents the perceptually dominant element during certain sections of a flamenco song.

The predominant melody extraction algorithm \cite{MELODIA} estimates pitch candidates on a frame level based on harmonic summation and groups them into pitch and time continuous contours using auditory streaming principles. It furthermore filters out contours based on their average pitch salience which do not form part of the dominant melodic line. As a result, it outputs a single pitch value $f_0[n]$ in Hz for all melody frames.
As this algorithm has previously been used in the context of flamenco singing, we adopt the parameters suggested in \cite{POLY} as follows: The analysis window corresponds to $N=4096$ samples with a hop size of $h_s=128$ samples at a sample rate of $f_s=44.1$kHz. The lower and upper limits for considered fundamental frequency candidates are set to 120Hz and 720Hz, respectively. These values correspond to the expected pitch range in flamenco singing. The voicing threshold $\tau_v [-2,3]$ which is related to the relative salience threshold for determining if a contour belongs to the main melody, is adjusted to $\tau_v=0.2$. It has been shown \cite{POLY} that with this value, around 90\% of all vocal frames are retained during the elimination process.  

\subsubsection{Contour filtering}
\label{ssec:Contour filtering}

\begin{figure}
\begin{minipage}[b]{1.0\linewidth}
  \centering
  \centerline{\includegraphics[width=7.3cm]{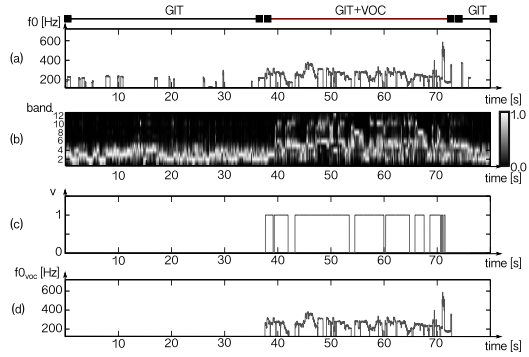}}
\end{minipage}
\vspace*{-0.3cm}
\caption{Contour filtering: (a) predominant melody, (b) lower bark bands, (c) vocal / non-vocal classifier and (d) estimated vocal melody.}
\label{fig:vocalDetection}
\end{figure}

We now apply a contour filtering process to decide which contours correspond to the vocal melody and to consequently eliminate guitar contours. The proposed scheme is based on two assumptions: First, the majority of all melody frames are voiced, or, in other words, the main melody largely corresponds to the vocal melody. This assumption is supported by the experiments carried out by G{\'o}mez {\it et al.}  \cite{POLY}, where only around 15\% of all estimated pitch values originate from the guitar. Second, we assume that guitar and vocal sections can be discriminated based on spectral characteristics. In preliminary experiments, we investigated a variety of spectral descriptors and found that the energy of the first 12 bark bands \cite{BARK} are best-suited for this task (Figure \ref{fig:vocalDetection} (a) and (b)). Based on these assumptions, we model melody frames originating from the guitar as outliers in a Gaussian distribution of bark band energies extracted for all melody frames. 

First, we extract the energy in the lower twelve bark bands with frequency limits specified as 0Hz, 50Hz, 100Hz, 150Hz, 200Hz, 300Hz, 400Hz, 510Hz, 630Hz, 770Hz, 920Hz, 1080Hz and 1270Hz in windows of length $N=1024$ samples with a hop size of $h_s=128$ samples at a sampling rate of $f_s=44.1$kHz. The energy in the $m^{th}$ bark band with lower frequency limit $f_{1,m}$ and upper frequency limit $f_{2,m}$ at time frame $n$ is given as 
\begin{equation}
\label{bark}
B[n,m]= {\sum\limits_{k(f_{1,m})<k<k( f_{2,m})}|{X}[k,n]|^2}
\end{equation}
where $k(f)$ is the frequency bin corresponding to the frequency $f$ (Eq. \ref{bins}) and $|{X}[k,n]|$ is the magnitude spectrum at time frame $n$. As a result, we obtain a feature vector $\mathbf{x}[n]=\langle B[n,1], ..., B[n,12] \rangle$ holding the energies of the lower twelve bark bands for each analysed frame $n$. 

We then assign an initial label to the feature vector in each frame based on the output of the predominant melody algorithm: Melody frames are labelled as {\em voiced} $\mathbf{x}_+$ and non-melody frames as {\em unvoiced} $\mathbf{x}_-$. This initial labelling corresponds to the case where the main melody coincides with the vocal melody. Subsequently, we fit a single multivariate Gaussian distribution to both feature sets separately. Applying maximum likelihood estimation, we obtain estimates for mean ($\boldsymbol{\mu}_+$ and $\boldsymbol{\mu}_-$) and covariance ($\mathbf{\Sigma}_+$ and $\mathbf{\Sigma}_-$) for both classes. The resulting likelihood $p$ for an arbitrary feature vector $\mathbf{x}$ is given as:
\begin{equation}
p(\mathbf{x} | \boldsymbol{\mu}, \mathbf{\Sigma})=\frac{1}{(2\pi)^{2}|\mathbf{\Sigma}|^{1/2}}\cdot e^{(-\frac{1}{2}(\mathbf{x}- \boldsymbol{\mu})^T \mathbf{\Sigma}^{-1}(\mathbf{x}- \boldsymbol{\mu}))}
\end{equation}
Evaluating this equation for a given feature vector $\mathbf{x}$ for both distributions, voiced $p_+(\mathbf{x} | \boldsymbol{\mu_+}, \mathbf{\Sigma_+})$ and unvoiced $p_-(\mathbf{x} | \boldsymbol{\mu_-}, \mathbf{\Sigma_-})$, we now expect a higher value for $p_+$ if the frame contains vocals. In this manner, we evaluate every frame and assign a binary vocal prediction $v[n]$:
\begin{equation}
\nu[n]=
\left\{
	\begin{array}{ll}
		1  & \mbox{if } p_+(\mathbf{x}[n] | \boldsymbol{\mu}_+, \mathbf{\Sigma_+}) \geq p_-(\mathbf{x}[n] | \boldsymbol{\mu}_-, \mathbf{\Sigma_-}) \\
		0 & \mbox{else} 
	\end{array}
\right.
\end{equation}
Since we assume voiced section to be continuous in time, we subsequently apply a binary moving average filter of 1 second length to eliminate fast fluctuations from the vocal prediction. The resulting sequence is shown in Figure \ref{fig:vocalDetection} (c). 

We now use this frame-wise classification to filter contour segments. A given contour ranging from frame $d_1$ to frame $d_2$ is eliminated, if it is entirely located outside the estimated vocal regions:

 \begin{equation}
\sum_{v[d_1]}^{v[d_2]}
\left\{
	\begin{array}{ll}
		==0  & \mathrm{eliminate} \\
		>0 & \mathrm{retain}\\
	\end{array}
\right.
\end{equation}
An example of the resulting vocal pitch is depicted in Figure \ref{fig:vocalDetection} (d). 

\subsection{Note transcription}
\label{ssec:note transcription}
After eliminating unwanted guitar contours, we are now left with a set of contours corresponding to the vocal melody. Instead of a continuous time-frequency representation, we now aim to segment the remaining contours into discrete note events, characterised by onset time, duration and a quantised semi-tone pitch value. In other words, the task is to split the contours at vocal onsets and assign a pitch label to each note event. 

\subsubsection{Segmentation}
\label{ssec:note segmentation}
Conceptually, we can distinguish two types of notes onsets: Those which coincide with a change in pitch ({\em interval onsets}) and onsets where the pitch of the current note is the same as the previous note ({\em steady pitch onsets}). In what follows, we will define four detection schemes which in their combination capture the majority of both types of onsets at a low false positive rate.

\begin{figure}
\begin{minipage}[b]{1.0\linewidth}
  \centering
  \centerline{\includegraphics[width=6.5cm]{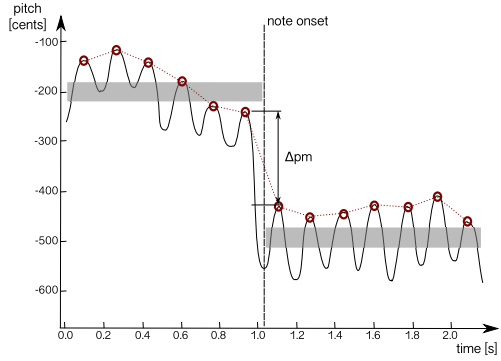}}
\end{minipage}
\vspace*{-0.3cm}
\caption{Note segmentation based on local maxima: Ground truth transcription (grey rectangles), pitch contour, local maxima (red circles) and upper envelope (red dashed line).}
\label{fig:locMax}
\end{figure}

For each voiced segment, we first map the frequency contour $f_0[n]$ to a cent scale relative to a reference frequency of $f_{\mathrm{Ref}} = 440$Hz:

\begin{equation}
\label{eq:map}
c_0[n]=1200*\log_2\left(\frac{f_0[n]}{f_{\mathrm{Ref}}}\right)
\end{equation}

First, we aim to discover {\em interval onsets}: When dealing with strongly ornamented pitch contours, the detection of significant pitch changes which indicate a note onset turns into a complex task. For the particular case of flamenco singing, fast fluctuations of the pitch track caused by micro-tonal ornamentations can exceed a semitone without a note onset being present. Low-pass filtering of the pitch contour reduces these fluctuations but also affects the steep slopes which indicate a note change. A close inspection of such pitch contours has shown that the relative change of the upper envelope gives a good indication of the underlying slowly changing perceived pitch (Figure \ref{fig:locMax}): Vocal vibrato causes a series of local maxima and in case of a steady note their pitch values remain in a strongly limited range in contrast to the actual pitch contour which might show fluctuations with an extend up to a semitone. In preliminary experiments the upper envelope has proven to be slightly more reliable for this task than the lower envelope. We furthermore observe that due to the periodicity of the vibrato fluctuations, the steep slope characterising a note onset is centred between two adjacent local maxima. We therefore extract the local maxima $pm$ and assume a pitch change centred between two adjacent maxima if their relative distance $\Delta \mathrm{pm}_{i,i+1}$ exceeds a threshold $\Delta p_{\mathrm{min}}$. For a pitch change of an exact semitone, we expect $\Delta \mathrm{pm}_{i,i+1}=100\mathrm{cents}$. In order to leave room for intonation inaccuracies, we adjust the threshold to $\Delta p_{\mathrm{min}}=80\mathrm{cents}$. Since not all local maxima are caused by vocal vibrato, we furthermore exclude cases where the time distance between two local maxima exceeds $T=0.25$s. We chose this values in order to cover modulation rates as low as $4$Hz, which corresponds to the lower bound of expected vocal vibrato rates \cite{VIBRATO}. With this procedure we detect a large number of pitch interval onsets, in particular in sections where vocal vibrato is present. Nevertheless, a number of onsets are missed, i.e. when the contour does not show a regular vibrato and consequently no relevant local maxima in the area of a pitch change. 

\begin{figure}
\begin{minipage}[b]{1.0\linewidth}
  \centering
  \centerline{\includegraphics[width=6.5cm]{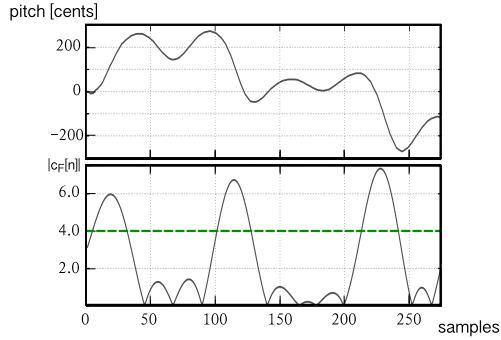}}
\end{minipage}
\vspace*{-0.3cm}
\caption{Gaussian derivative filtering; top: cent-scaled pitch contour; bottom: output of the Gaussian derivative filter. The green dashed line marks the detection threshold for local maxima.}
\label{fig:GF}
\end{figure}

In order to detect further undiscovered onsets, we refine the segmentation by applying a first derivative Gaussian filter to the contour. Such filters have been used successfully in the related task of edge detection in image processing \cite{EDGE1,EDGE2}. The one-dimensional Gaussian density function $g[n, \sigma]$ with zero mean and standard deviation $\sigma$ is given as:
\begin{equation}
g[n, \sigma]=e^{-\frac{n^2}{2\sigma^2}}
\end{equation}
and its first derivative h[n] results to:
\begin{equation}
h[n]=\frac{\delta g[n, \sigma]}{\delta n}=-\frac{n}{\sigma^2} \cdot  e^{-\frac{n^2}{2\sigma^2}}
\end{equation}
In the context of onset detection, the purpose of applying Gaussian derivative filtering is to detect long-term changes in the signal while omitting fast fluctuations. The parameter $\sigma$ determines the effective length of the filter and consequently the period of averaging: If $\sigma$ is too large, the filter might average entire short notes. Choosing a very small value for $\sigma$ will lead to a large filter output at fast pitch fluctuations caused by vibrato. We choose $\sigma=43.5$ms, so that the effective filter length of $300$ms covers a full period of a slow vibrato with a rate of $4$Hz. Applying the filter $h[n]$ of length $N_h$ to the cent-scaled contour $c_0[n]$ of length $N_c$ leads to the filtered output $c_F[n]$. Analysing its absolute value (Figure \ref{fig:GF}) shows strong peaks in the area where a change of the slowly varying underlying pitch takes place. We consequently detect onsets at local maxima of $|c_F[n]|$. Since minor variations in the pitch contour may cause noise in the filter output, we discard local maxima below an empirically determined threshold of $|c_F[n]|_{\mathrm{min}}=4.0$. 

\begin{figure}
\begin{minipage}[b]{1.0\linewidth}
  \centering
  \centerline{\includegraphics[width=7.3cm]{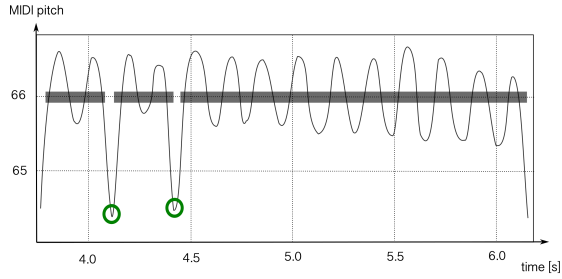}}
\end{minipage}
\vspace*{-0.3cm}
\caption{Pitch contour dips at note onsets: pitch contour; note onsets characterised by a dip in the pitch contour. Green circle mark detected onsets, grey rectangles correspond to the ground truth transcription.}
\label{fig:rms}
\end{figure}

\begin{figure}
\begin{minipage}[b]{1.0\linewidth}
  \centering
  \centerline{\includegraphics[width=7.3cm]{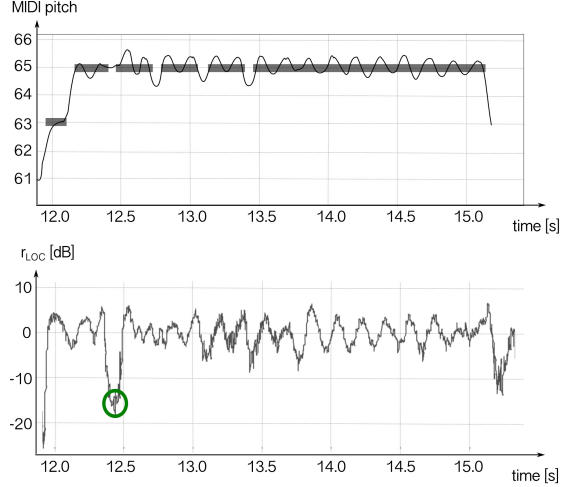}}
\end{minipage}
\vspace*{-0.3cm}
\caption{Note onsets characterised by a local decrease in volume; top: pitch contour; bottom: local RMS; The green circle marks a detected onset, grey rectangles correspond to the ground truth transcription.}
\label{fig:dip}
\end{figure}

We now proceed to the detection of {\em steady pitch onsets}: We define two characteristics which indicate this type of onset: A sudden local decay in volume (Figure \ref{fig:rms}) and sudden decrease in the pitch contour (Figure \ref{fig:dip}). It is important to state at this point, that at a given onset either one or both of these characteristics can be present. We therefore define two separate detection schemes instead of combining both features in a single detection function. Detecting volume decays when the overall dynamics of a track vary, i.e. certain sections are generally of lower volume than others, requires an analysis of the short-term dynamics. We first extract the root mean square (RMS) of the signal $\mathrm{rms}[n]$ in windows of length $N=4096$ samples with a hop size of $h_S=128$ samples. In order to detect local decays, we define the local RMS fluctuation $r_{\mathrm{LOC}}[n]$ by comparing each sample to the mean value of its surrounding $100$ samples, corresponding to $150$ms and map to a decibel scale:
\begin{equation}
r_{\mathrm{LOC}}[n]=20*\log10\left(\frac{\mathrm{rms}[n]}{\sum_{n-50}^{n+50}{\mathrm{rms}[n]}*0.01}\right)
\end{equation}
We segment the contour at frames $n$ where the local minima with $r_{\mathrm{LOC}}[n]<-10$dB are found. We set this threshold in order to ignore local volume variations which often accompany pitch vibrato or are caused by dynamic fluctuations of the accompaniment. 

The difficulty in detecting the previously described decreases in the pitch envelope is to distinguish them from local minima occurring during vocal vibrato. We therefore model local minima which are related to note onsets as outliers in the local distribution of pitch values over all frames in the considered contour. A common method to detect outlier values of a distribution is to analyse its z-score $z[n]$, which describes the relation between a given datapoint and the standard deviation of all considered data points,
\begin{equation}
z[n]=\frac{c[n]-\mu}{\sigma}
\end{equation}
where $\mu$ denotes the mean and $\sigma$ the standard deviation of the distribution of pitch values in the considered segment. Consequently, the described decreases in a given contour will produce large negative values for $z[n]$. In order to avoid detecting local minima caused by vocal vibrato, we only consider negative peaks of $z[n]$ below a threshold $z_{\mathrm{max}}$. The standard deviation of a zero-centred sinusoid $\sigma_{\mathrm{sin}}$ with a magnitude $A$ corresponds to its RMS value and results to:
\begin{equation}
\sigma_{sin}=\frac{A}{\sqrt{2}}
\end{equation}
Local minima of the periodic oscillation will consequently cause a z-score $z_{\mathrm{min,sin}}$ of
\begin{equation}
z_{\mathrm{min,sin}}=\frac{-A}{\sigma_{\mathrm{sin}}}=-\sqrt{2}
\end{equation}
Since we are looking for local pitch decreases which ares significantly larger than the deviations caused by vocal vibrato and furthermore the vibrato extend might vary during long notes, we adjust the threshold for detecting local minima below the theoretical boundary of $-\sqrt{2}$ to $z_{\mathrm{max}}=-2$.

\subsubsection{Pitch labelling}
\label{ssec:Pitch labeling}
\begin{figure}
\begin{minipage}[b]{1.0\linewidth}
  \centering
  \centerline{\includegraphics[width=7.3cm]{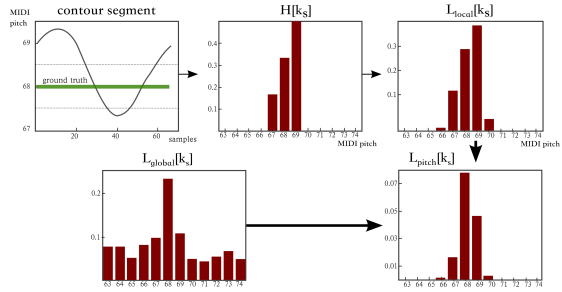}}
\end{minipage}
\vspace*{-0.3cm}
\caption{Pitch labelling: (a) contour segment; (b) local pitch histogram; (c) local pitch probability; (d) global pitch probability; (e) combined pitch probability.}
\label{fig:label}
\end{figure}
After the note segmentation stage, the remaining task is to assign a pitch label to each resulting note event. There are various challenges in this stage: First, the tuning of the track might deviate from the reference, causing even constant segments to be located between two semitone bins. While the mean pitch value across the contour is often a sufficient approximation for stable contour segments or even symmetric ornamentations (i.e. vibrato), singing voice contours may contain portamentos at note beginnings and endings or over-swing before the target pitch is reached. Such non-symmetric ornamentations can cause an offset in the mean pitch value. Finally, local intonation might not always be accurate and the estimated pitch can be located slightly below or above the target value. McNab \cite{MCNAB97} estimates the resulting pitch as the peak bin of the local pitch histogram. Molina {\it et al.} determine the pitch label of a given segment as the energy-weighted alpha-trimmed mean in order to exclude local pitch outliers and give assign higher weights to high-energy frames. In the model-based transcription system \cite{RY06}, the pitch label of each found note results directly from the output of the HMM. In a similar way, G{\'o}mez \& Bonada \cite{MONO} deduce quantised pitch labels from the maximum likelihood estimation. 

Here, we first estimate a global tuning reference for the entire song and then combine pitch statistics from the note event with global pitch class probabilities to finally determine the pitch value. The entire process is summarised in Figure \ref{fig:label}.

We estimate a global tuning deviation $\Delta t$ from the reference value of $A_{4,\mathrm{St}}=440$Hz using an established method based on circular statistics \cite{TUNING}. The new tuning reference $A_{4,T}$ results to 
\begin{equation}
A_{4,T}=2^{\frac{\Delta t}{1200}} \cdot A_{4,\mathrm{St}}
\end{equation}
and we can re-map each contour $f_0[n]$ to a cent scale $c_{0,T}[n]$ relative to the estimated reference frequency $f_{\mathrm{Ref}}=A_{4,T}$ applying Eq. \ref{eq:map}.

Even though in several music traditions and genres, the main melody does not strictly follow a defined scale, certain pitches tend to occur more frequent than others according to the underlying harmonic context. For note events with unstable pitch or inaccurate intonation, such probabilities can assist in deciding on the pitch label. We estimate a so called pitch class profile providing probability estimates for all twelve semitones from the averaged chroma vector over all frames. Chroma features were first introduced by Wakefield \cite{CHROMA} and are frequently used in the context of a variety of MIR tasks, such chord and tonality estimation or cover song identification \cite{CHROMA2}. The chroma vector for a given time frame is obtained by quantising the spectrum $|X[n,k]|$ to semi-tone bins and then mapping the entire analysed range into a single octave. In this manner, we obtain an instantaneous chroma vector $\mathrm{chr}[k_{\mathrm{chr}},n]$ for each frame $n$ consisting of 12 semitone bins $k_{\mathrm{chr}}$. Subsequently, we can estimate a global pitch class probability $L_{\mathrm{global}}$ for each semitone from the average chroma vector $\overline{\mathrm{chr}[k_{\mathrm{chr}}]}$ over all frames:
\begin{equation}
L_{\mathrm{global}}[k_{\mathrm{chr}}]=\frac{\overline{\mathrm{chr}[k_{\mathrm{chr}}]}}{\sum\limits_{k_{\mathrm{chr}}=1}^{12}\overline{\mathrm{chr}[k_{\mathrm{chr}}]}}
\end{equation}
The use of chroma vectors give the advantage that it includes pitch information of both, the singing voice and the guitar accompaniment, and should therefore give a better representation of the underlying tonality.

We now proceed to the statistical analysis of the pitch content within a single note events. The centre values $C_c[k_s]$ in cents of quantised cent bins corresponding to the $k_s^{th}$ semitone above or the $-k_s^{th}$ semitone below $A_{4,T}$ are given as:
\begin{equation}
C_c[k_s]=k_s \cdot 100
\end{equation}
Now, for each frame within the note event, the pitch contour $c_{0,T}[n]$ is quantised to the closest semitone bin $k_s$ and accordingly a pitch histogram $H[k_s]$ is computed by accumulating the occurrences of each bin and dividing by the total number of frames. As mentioned earlier, due to local intonation inaccuracies and non-symmetric ornamentations, the ground truth pitch might not correspond to the peak bin but to one of the adjacent bins. We therefore replace each bin value $H[k_s]$ by a Gaussian distribution $G_{k_s'}[k_s]$ originating from the bin $k_s'$ and spanning over various semitone bins $k_s$: 
\begin{equation}
G_{k_s'}[k_s]=H[k_s'] \cdot \frac{1}{\sigma \sqrt{2 \pi}} e^{-\frac{(k_s-k_s')^2}{2 \sigma^2}}
\end{equation}
The mean of the distribution corresponds to the semitone bin $k_s'$ from which the distribution originates. We furthermore assume a standard deviation of $\sigma=0.5$ corresponding to a quarter tone and weight with the occurrence of the respective bin $H[k_s']$. By accumulating the contributions of all distributions for each bin, we obtain the local pitch probability function $L_{local}[k_s]$ as a Gaussian mixture distribution:
 \begin{equation}
 L_{\mathrm{local}}[k_s]=\sum_{k_s'} G_{k_s'}[k_s]
 \end{equation}

The pitch label can now be estimated by combining the global pitch class probability and the local pitch probability function. For a given semitone bin $k_s$ the corresponding chroma bin $k_{chr}$ is defined as: 
 \begin{equation}
 k_{\mathrm{chr}}[k_s]=k_s\mod 12
 \end{equation} 
 The resulting pitch probability for a semitone $L_{\mathrm{pitch}}[k_s]$ is calculated as the product of global and local pitch probabilities:
 \begin{equation}
 L_{\mathrm{pitch}}[k_s]=L_{\mathrm{global}}[k_{\mathrm{chr}}[k_s]] \cdot L_{\mathrm{local}}[k_s]
 \end{equation}
The MIDI pitch label $P_{\mathrm{midi}, i}$ associated with the $i^{th}$ contour segment is finally computed from the semitone bin with the highest combined pitch probability:
\begin{equation}
P_{\mathrm{midi},i}=69+\underset{k_s} {\mathrm{argmax}} ~(L_{\mathrm{pitch},i}[k_s])
\end{equation} 

\subsubsection{Note post-processing}
\label{sec:thinning}
\begin{figure}
\begin{minipage}[b]{1.0\linewidth}
  \centering
  \centerline{\includegraphics[width=6.3cm]{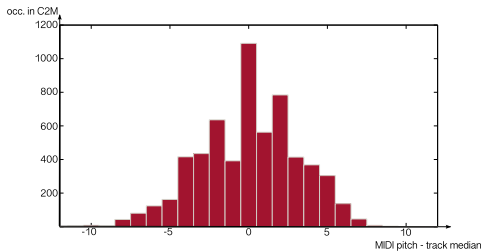}}
\end{minipage}
\vspace*{-0.3cm}
\caption{Distribution of ground truth pitch values with respect to the track median for the {\em cante2midi} dataset.}
\label{fig:pitchRange}
\end{figure}

By applying basic musicologically motivated restrictions regarding pitch and duration, we can further refine the obtained note transcription. First, we can exploit the idea that flamenco singing is usually limited to a pitch range of less than an octave. Figure \ref{fig:pitchRange} shows an analysis of the {\em cante2midi} dataset described in Section \ref{sec:evaluation}, where the pitch of a ground truth note is displayed in relation to the median pitch of the corresponding track. Based on this observations, we subtract the median pitch of the entire track from each transcribed note and limit the range to $\pm8$ semitones. Transcribed notes below this range are likely to belong to the accompaniment and are consequently eliminated. Notes above this range may be caused by octave errors in the vocal melody extraction and are transposed down by one octave. We furthermore assume a minimum note duration of $0.05$ seconds and eliminate shorter transcribed notes. In the analysed datasets (\ref{ssec:testCollection}), only 0.3\% of all ground truth notes have a shorter duration. In order to reduce computational complexity and to avoid labelling short notes which are afterwards discarded, segments shorter than $0.05$ seconds are eliminated before the note labelling stage. 

\section{Evaluation Methodology}
\label{sec:evaluation}
Below we provide a detailed description of the evaluation strategies applied in the scope of this study. We give an overview of the employed data collections and the corresponding ground truth annotation process and describe the evaluation metrics used. We furthermore provide a short description of the reference algorithms used during the comparative evaluation.  

\subsection{Test collections}
\label{ssec:testCollection}
We use three data collections to evaluate the performance of the proposed approach and compare to reference algorithms: The {\em cante2midi} (C2M) dataset was gathered in the scope of this study and contains a variety of flamenco styles and voice timbres with varying degree and complexity of ornamentation. The instrumentation of all 20 tracks is limited to vocals and guitar. The {\em fandango} (FAN) dataset comprises 39 excerpts from recordings of the fandango style, containing vocals and guitar accompaniment. This dataset was used previously in the context of automatic transcription of flamenco singing \cite{POLY}. In order to compare to monophonic transcription systems, we furthermore evaluated the proposed system on the {\em tonas} (TON) dataset. This collection contains 72 clips of a cappella flamenco recordings and is publicly available\footnote{mtg.upf.edu/download/datasets/tonas/}. It should be mentioned that a cappella singing represents only a small fraction of the flamenco genre and such performances are usually characterised by a large amount of melismatic ornamentation. Furthermore, the absence of the guitar accompaniment often causes strong tuning fluctuations throughout the performance. Further information on all three data collections is provided in Table \ref{tab:data}. 

\begin{table}
\small
\begin{center}
\renewcommand{\arraystretch}{1.2}
\begin{tabular}{lccc}
\hline
\textbf{database}    & \textbf{cante2midi} & \textbf{fandangos} & \textbf{tonas} \\
\hline
no. tracks  & 20 & 39 & 72 \\
clip type  & full track & excerpt & excerpt \\
no. singers  & 15 & 21 & 44 \\
total duration & 1h 6m & 34m & 20m \\
no. ground truth notes & 6025 & 3070 & 2983 \\
\% voiced frames  & 42 & 50 & 82 \\
\hline
\end{tabular}
\end{center}
%\vspace*{-0.3cm}
\caption{Database information.}
\label{tab:data}
\end{table}

\subsection{Ground truth annotations}
\label{ssec:groundTruth}
A general guideline for the ground truth annotation process of all three collections was to obtain a detailed transcription including all audible notes. Since flamenco music does not always follow a strict rhythm and the proposed transcription system does not apply rhythmic quantisation, note onsets were transcribed as absolute time instants. The ground truth annotations for the C2M collection were conducted by a person with formal music education and training in melody transcription by ear, but only basic knowledge of flamenco. All transcriptions were verified and corrected by a flamenco expert. The output of the system described by G{\'o}mez {\it et al.} \cite{POLY} was taken as a starting point. After converting to MIDI, transcriptions were edited in the digital audio workstation {\em LogicPro}. The annotator listened to the original track and the transcription synthesised by a piano simultaneously with the possibility of muting one of the tracks when required. The tuning of the MIDI synth was manually adjusted to match the tuning of the corresponding audio track. A visual representation of the pitch contour and the baseline transcription was provided as additional aid. Ground truth annotations for both, the FAN and the TON collection, were conducted by a musician with limited knowledge of flamenco in order to avoid implicit knowledge of the style. Annotations were then verified and corrected by a flamenco expert and occasionally discussed with another flamenco expert. For more details on the annotation process of these two collections and general guidelines of manual flamenco singing transcription, we refer to \cite{MONO} and \cite{POLY}. In addition to the note transcriptions, we furthermore provide manually corrected pitch contours for both polyphonic databases, where guitar contours were eliminated. 

\subsection{Reference algorithms}
\label{ssec:referenceAlgorithms}
In the course of this study we compared the proposed system to a number of reference algorithms. Below we briefly describe each of the methods we used in the comparative evaluation in Section \ref{sec:experiments}. For a detailed description we refer the reader to the references provided for each method. 

\begin{itemize}

\item \textit{Curve fitting (FIT)}: A contour simplification algorithm \cite{FITTING} was applied to a given pitch contour. As suggested by the authors, the pitch deviation tolerance was set to $\alpha_e$=1. 
 
\item \textit{Recursive least squares filtering (RLS)}:
This approach performs a segmentation of the pitch contour based on the error function of an adaptive filter which tracks the semitone pitch contour as described by Adams \cite{ADAMS06}. We adopt all system parameters suggested by the authors and segment at values of the squared error function $d[n]>0.25$. 

\item \textit{Dynamic programming (DP)}:
We used an existing implementation of the flamenco singing transcription system described by G{\'o}mez \& Bonada \cite{MONO} (DP-Mono) to transcribe a cappella recordings and applied its extension  \cite{POLY} (DP-Poly) for polyphonic recordings. The implementation furthermore allows to segment any given pitch sequence.

\item \textit{Segmentation based on hysteresis (SiPTH)}:
The monophonic singing transcription described by Molina {\it et al.} \cite{SIPTH} was used to transcribe a cappella singing recordings.

\item \textit{Segmentation based on probabilistic Hidden Markov Modelling (HMM)}:
We compare to the implementation of the segmentation algorithm proposed in \cite{RY06} as implemented in the {\em tony} \cite{TONY} transcription framework for monophonic singing recordings. The system uses the probabilistic YIN \cite{PYIN} algorithm as a front end to estimate the vocal pitch. We used the publicly available vamp plugin implementation\footnote{code.soundsoftware.ac.uk/projects/pyin}.  
\end{itemize}

The systems SiPTH and HMM represent monophonic singing transcription frameworks and were consequently evaluated in the scope of a cappella singing. FIT and RLS are contour segmentation algorithms without a vocal pitch extraction front end and were used in the scope of onset detection evaluation. The DP algorithm is the only complete reference system for singing transcription from polyphonic recordings and was therefore part of a comparative study evaluating the overall system performance. Since it can furthermore operate on any given pitch contour input, it was also used in the onset detection evaluation. 

\begin{figure}
\begin{minipage}[b]{1.0\linewidth}
  \centering
  \centerline{\includegraphics[width=5.0cm]{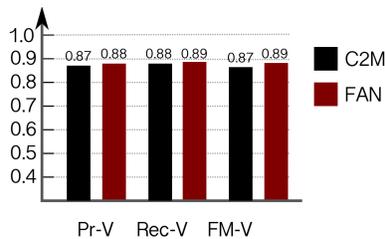}}
\end{minipage}
\vspace*{-0.3cm}
\caption{Frame-wise accuracy of the GMM-based vocal detector \cite{VD1}: {\em cante2midi} and {\em fandango} dataset.}
\label{fig:GMM}
\end{figure}

\begin{itemize}
\item \textit{Vocal detection using Gaussion Mixture Models (GMM)}:
In order to evaluate the proposed vocal detection stage to existing methods, we furthermore implemented a the algorithm proposed by Song {\it et al.} \cite{VD1} based on Gaussian Mixture Models. We adopted all parameters as suggested by the authors and processed both polyphonic datasets in a 10-fold cross-validation. The frame-wise accuracy by means of voicing precision, recall and f-measure is shown in \ref{fig:GMM} and is slightly above the values reported by the authors
\end{itemize}

\subsection{Evaluation metrics}
\label{sec:evaluationMetrics}
We aim to evaluate three aspects of the proposed algorithm: Its capability of detecting the vocal sections, the performance of the segmentation of vocal contours into discrete note events and its overall performance. 

The vocal section retrieval is evaluated by means of voicing precision (Pr-V), voicing recall (Rec-V) and voicing f-measure (FM-V). Pr-V is defined as the fraction of all frames estimated as voiced, which are labelled as voiced in the ground truth. Rec-V corresponds to the fraction of all voiced ground truth frames, which are estimated as voiced. The resulting f-measure is calculated as the harmonic mean of Rec-V and Pr-V.

In a similar manner, we evaluate the onset detection stage by means of onset precision (Pr-On),  onset recall (Rec-On) and onset f-measure (FM-On). In this case, Pr-On refers to the proportion of all detected onsets, which correspond to ground truth onsets. Rec-On is defined as the proportion of all ground truth onsets, which are correctly detected. FM-On again corresponds to the harmonic mean of Rec-On and Pr-On. We adapt a previously suggested threshold (\cite{MONO}) and consider an onset as correctly detected if it is located within $0.15$ seconds of a ground truth onset. Furthermore, each ground truth onset can only be associated to a single detected onset and vice versa. 

In order to evaluate note transcriptions, we first define the frame-wise raw pitch accuracy (RPA) as the percentage of correctly transcribed frames. In order to incorporate the voicing detection into this measure, we define a an unvoiced frame as correctly transcribed if it was estimated as unvoiced. Voiced frames are correctly transcribed if the estimated MIDI pitch corresponds to the ground truth in this frame. We furthermore evaluate transcriptions based on note precision (Pr-N), note recall (Rec-N) and note f-measure (FM-N). Pr-N refers to the proportion of all detected notes, which are correctly transcribed ground truth notes. Rec-N is defined as the proportion of all ground truth notes, which are correctly transcribed and FM-N is calculated as the harmonic mean of Pr-N and Rec-N.

We adopt the evaluation thresholds suggested in \cite{MONO}: A ground truth onset is correctly detected if an estimated onset is within a range of $0.15$ seconds. Furthermore, a note is correctly transcribed if the pitch label is correctly assigned and the estimated duration is within a tolerance of $30\%$. Analysing the test collection, we observe significant deviations from the standard tuning of $A_4=440$Hz ranging even above $40$ cents. For such large deviations, it is even in the manual process difficult to decide, if the track is tuned below or above the reference. Consequently, small errors in the tuning estimation of only a few cents can cause the entire melody to be transcribed a semitone above or below the ground truth. We assume that in cases of large tuning deviations such a transcription should still be valid. Therefore, we perform a preliminary evaluation of the entire transcription and its transpositions one semitone above and below and correct towards the best match.

\section{Experimental Results}
\label{sec:experiments}
In this section, we present the results of a number of experiments carried out in order to evaluate the performance of the proposed system (P) and compare to other methods. We first evaluate the overall performance for flamenco transcription from polyphonic and monophonic systems and compare to a several existing systems. In order to deal with monophonic recordings, we omit the channel selection and contour filtering stage and increase the voicing tolerance of the predominant melody extraction algorithm (Section \ref{ssec:Predominant melody extraction}) to $\tau_v=3.0$. This setup is denoted as P-Mono. We then analyse the accuracy of the proposed contour filtering stage, compare to alternative system setups and investigate the influence on the note transcription. Subsequently, we first isolate the contour segmentation stage and then the entire note transcription block and study the obtained accuracies with respect to alternative approaches. Finally, we conduct a component analysis of the proposed system and investigate the influence on the performance when certain processing blocks are removed.

\subsection{Overall system}
\label{ssec:overall}
\begin{figure}
\begin{minipage}[b]{1.0\linewidth}
  \centering
  \centerline{\includegraphics[width=6.8cm]{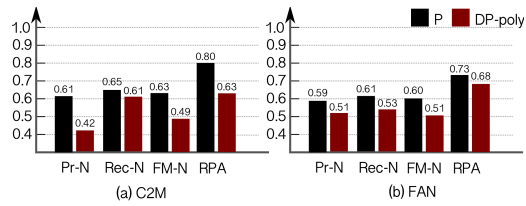}}
\end{minipage}
\vspace*{-0.3cm}
\caption{Overall system evaluation for polyphonic datasets: Proposed system (P) and \cite{POLY} (DP-Poly).}
\label{fig:R_overall_poly}
\end{figure}

\begin{figure}
\begin{minipage}[b]{1.0\linewidth}
  \centering
  \centerline{\includegraphics[width=7.3cm]{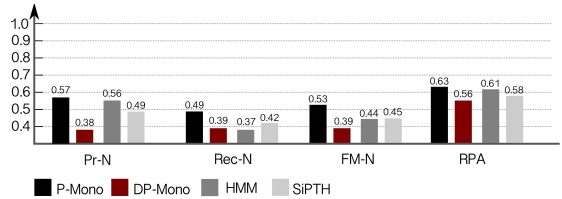}}
\end{minipage}
\vspace*{-0.3cm}
\caption{Overall system evaluation for the monophonic dataset (TON): Proposed system (P), \cite{MONO} (DP-Mono), \cite{RY06, TONY} (HMM) and \cite{SIPTH} (SiPTH).}
\label{fig:R_overall_tonas}
\end{figure}

Figure \ref{fig:R_overall_poly} shows the evaluation of the proposed system (P) and the polyphonic implementation of the dynamic programming approach (DP-Poly) \cite{POLY} for the two polyphonic datasets described in Section \ref{ssec:referenceAlgorithms}. Figure \ref{fig:R_overall_tonas} shows the comparative evaluation on the monophonic dataset. For both, monophonic and polyphonic recordings, the proposed system outperforms all reference systems. The note f-measure as well as the frame-wise raw pitch accuracy is significantly lower for the monophonic dataset, indicating the difficulty of transcribing this particular sub-genre (Section \ref{ssec:testCollection}). 

\subsection{Voicing}
\label{ssec:R_voicing}
\begin{figure}
\begin{minipage}[b]{1.0\linewidth}
  \centering
  \centerline{\includegraphics[width=6.8cm]{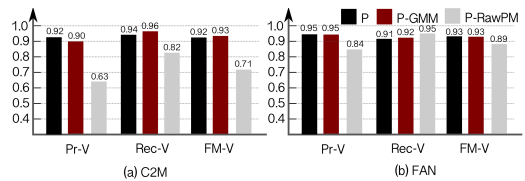}}
\end{minipage}
\vspace*{-0.3cm}
\caption{Frame-wise voicing accuracy for the proposed system (P), the proposed system with \cite{VD1} replacing the vocal detection function $v[n]$ (P-GMM) and the raw predominant melody (P-RawPM).}
\label{fig:R_voicing_V}
\end{figure}

\begin{figure}
\begin{minipage}[b]{1.0\linewidth}
  \centering
  \centerline{\includegraphics[width=7.3cm]{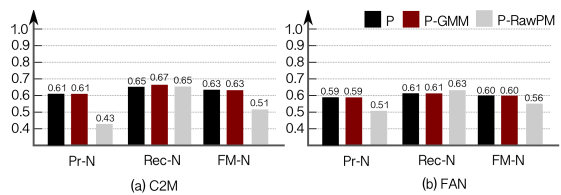}}
\end{minipage}
\vspace*{-0.3cm}
\caption{Note transcription evaluation for the proposed system (P), the proposed system with \cite{VD1} replacing the vocal detection function $v[n]$ (P-GMM) and the raw predominant melody (P-RawPM).}
\label{fig:R_voicing_N}
\end{figure}

We now investigate the effectiveness of the proposed vocal pitch extraction stage (P) and compare to two alternative setups: P-RawPM refers to replacing the entire stage with the raw predominant melody without any further processing. This setup corresponds to the front-end of the algorithm proposed by G{\'o}mez \& Bonada \cite{POLY} and represents the baseline. We furthermore replace the proposed frame-wise voicing prediction $v[n]$ (Section \ref{ssec:Contour filtering}) with the output of the GMM-based approach described in \cite{VD1} and eliminate contours accordingly. This setup is referred to as P-GMM. The experiments were conducted for both polyphonic datasets, evaluating the frame-wise voicing accuracy (Figure \ref{fig:R_voicing_V}) as well as the influence on the resulting note transcription (Figure \ref{fig:R_voicing_N}). 

The results show that the raw predominant melody gives significantly better results for the {\em fandango} (FAN) than for the {\em cante2midi} (CSM) dataset. An explanation for this behaviour can be found when analysing the content of the excerpts comprising FAN: Despite containing only 50\% of voiced frames (Table \ref{tab:data}), the excerpts do not include the guitar introduction or melodic interludes. In contrast, C2M contains full tracks where these sections are included. These parts of the song represent the sections, where the guitar is most dominant and consequently tends to produce melody contours. Nevertheless, applying channel selection and contour filtering leads to a significant increase in the obtained frame-wise voicing accuracy which propagates to the resulting transcription quality: The proposed approach yields a note f-measure of $0.63$ on the C2M dataset and $0.60$ on the FAN dataset. The note f-measure obtained with the raw predominant decreases $0.51$ for C2M and $0.56$ for FAN. We can furthermore observe, that replacing $v[n]$ with the output of the GMM-based vocal detection yields similar results when compared to the proposed system. The difference in voicing accuracy between the two setups is marginal and does not show in the resulting not transcription performance (both obtain an f-measure of $0.63$ on C2M and $0.60$ on FAN). Nevertheless, an advantage of the proposed system is that it works on a track-level and does not require any training phase involving manual ground truth annotations.

\subsection{Segmentation}
\begin{figure}
\begin{minipage}[b]{1.0\linewidth}
  \centering
  \centerline{\includegraphics[width=5.5cm]{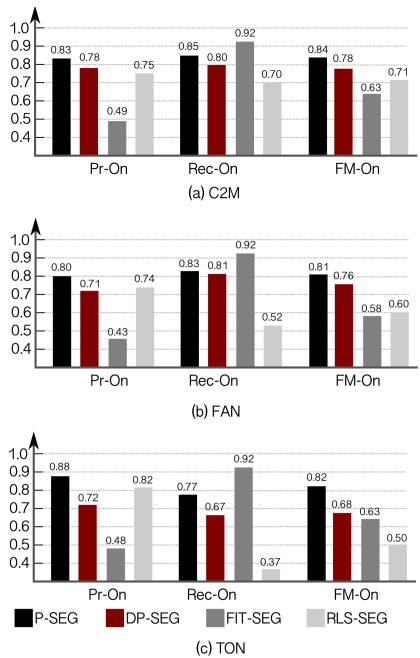}}
\end{minipage}
\vspace*{-0.3cm}
\caption{Note segmentation evaluation for the proposed system (P-SEG), the dynamic programming approach (DP-SEG) \cite{MONO}, the fitting algorithm (FIT-SEG) \cite{FITTING} and the RLS segmenter (RLS) \cite{ADAMS06}.}
\label{fig:R_onsets}
\end{figure}

We now isolate the note segmentation stage described in Section \ref{ssec:note segmentation} and compare to a number of alternative approaches. The evaluation is carried out by means of onset precision, recall and f-measure. For all three databases, the corrected vocal pitch contour is provided as input to all methods. The results shown in Figure \ref{fig:R_onsets} show that the proposed approach (P-SEG) yields the best performance among all considered methods on all datasets, followed by the dynamic programming approach (DP-SEG) \cite{MONO}. The curve fitting algorithm (FIT-SEG) \cite{FITTING} obtains a low precision but a high recall rate, indicating an over-segmentation of the contour. The reversed behaviour is observed for the RLS segmented (RLS-SEG) \cite{ADAMS06}. We furthermore observe, that  the performance of the proposed approach is consistent for the three dataset. The onset f-measure of DP-SEG decreases for the monophonic dataset ($0.78$ for C2M, $0.76$ for FAN and $0.68$ for TON). Given the higher complexity of the segmentation task for the {\em tonas} dataset described in Section \ref{ssec:testCollection}, these results indicate that the proposed approach is robust towards tuning inaccuracies and is capable of dealing with a large amount of melismatic ornamentations of the melody. 

\subsection{Note transcription}
\begin{figure}
\begin{minipage}[b]{1.0\linewidth}
  \centering
  \centerline{\includegraphics[width=6.5cm]{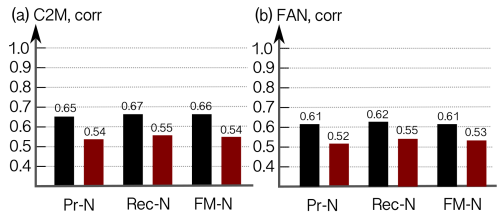}}
\end{minipage}
\vspace*{-0.3cm}
\caption{Note transcription evaluation for the proposed system (P), the dynamic programming approach (DP) \cite{MONO} when applied to the manually corrected pitch contour ({\em corr}).}
\label{fig:R_overall_corr}
\end{figure}

After isolating the note segmentation stage, we now proceed to the evaluation of the entire note transcription stage (Section \ref{ssec:note transcription}), comprising contour segmentation, pitch labelling and note post-processing. We compare to the DP algorithm \cite{MONO} and provide both systems with the manually corrected vocal pitch contour as input. This experimental setup represents a glass ceiling evaluation in a sense that it corresponds to a transcription with a perfect vocal pitch extraction stage in terms of voicing. Figure \ref{fig:R_overall_corr} provides the note-related evaluation for both polyphonic datasets. It can be observed that the decrease in performance from using the corrected pitch contours in this experiment to the real-world scenario (Figure \ref{fig:R_overall_poly}) is significantly lower for the proposed system: For the C2M dataset, the note f-measure drops from $0.66$ to $0.63$ for the proposed system and from $0.54$ to $0.39$ for the DP algorithm. This indicates that the proposed system provides a better approximation of the vocal pitch contour than the raw predominant melody \cite{MONO}. These results confirm the findings in Section \ref{ssec:R_voicing}. 

\subsection{Component analysis}
In a last experiment, we conduct a component analysis of the proposed system in order to verify that each of the core algorithm components contribute to the overall system performance. For this analysis, we evaluate the resulting note transcriptions for the C2M dataset for four system setups:

\begin{itemize}
\item P: proposed system
\item P-CF: proposed system without the contour filtering stage (Section \ref{ssec:Contour filtering}).
\item P-CS: proposed system without previous channel selection (Section \ref{ssec:channel selection}).
\item P-PP: proposed system without estimation of the global pitch probability (Section \ref{ssec:Pitch labeling}). Pitch labels are assigned based on local pitch histograms only. 
\end{itemize}

\begin{figure}
\begin{minipage}[b]{1.0\linewidth}
  \centering
  \centerline{\includegraphics[width=6.5cm]{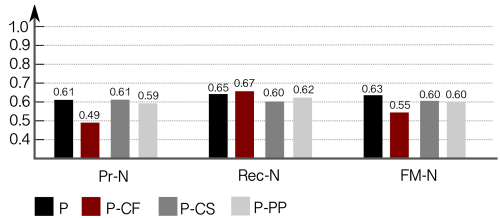}}
\end{minipage}
\vspace*{-0.3cm}
\caption{Note transcription evaluation for the proposed system (P) when several components are removed: Contour filtering (P-CF), channel selection (P-CS) and global pitch probability estimation (P-PP).}
\label{fig:R_components}
\end{figure}

The results displayed in Figure \ref{fig:R_components} confirm that each of the investigated algorithm components contributes to the overall system performance. Among the considered setups, the removal of the contour filtering stage caused the largest decrease in performance, reducing the note f-measure from $0.63$ to $0.55$. Nevertheless, this result is still superior to the performance observed for the system presented in \cite{POLY} (f-measure $0.49$).

\section{Conclusions}
\label{sec:conclusions}
In this paper we present a novel approach for automatic transcription of flamenco singing directly from polyphonic audio recordings. All involved signal processing blocks have been described in detail: For stereo recordings we first select the channel with stronger dominance of the vocals for further processing. We then extract the predominant melody and apply a novel contour filtering process to discard guitar contours. It has been shown that this process significantly improves the voicing accuracy when compared to the raw predominant pitch contour. The resulting estimated vocal pitch sequence is further processed in the note transcription stage, where contours are converted to discrete note events. We propose a novel contour segmentation procedure based on pitch and volume characteristics, which has proven to yield better results by means of onset detection accuracy when compared to a number of alternative approaches. We assign a pitch label to each resulting segment by combining local and global pitch information. A number of experiments have shown that the overall system as well as isolated stages outperform various reference algorithms. Our approach has proven to give convincing results given with the particular characteristics of flamenco singing, in particular strong ornamentations, extensive use of vocal vibrato and local intonation inaccuracies. The resulting automatic transcriptions therefor provide a suitable basis for a number of related MIR tasks, such as melodic similarity characterisation or automatic style identification. The system can furthermore aid in large-scale musicological studies by providing first estimates for computer-assisted transcription. 

There are several aspects in the context of flamenco singing transcriptions which we aim to investigate in order to further improve the quality of automatic transcription. A main topic of interest is to include perceptual aspects in the pitch labelling stage, by exploring the influence of fast pitch fluctuations and guitar accompaniment on the perceived pitch. Furthermore, it would be interesting to investigate how the quantitative measures presented in this study relate to perceptual transcription quality and how the transcription accuracy influences the performance of MIR systems which rely on automatic transcriptions. We are furthermore interested in how far more detailed note representations, i.e. including micro-tonal pitch labels compare the proposed standard MIDI representation for different MIR tasks. Finally, we aim to annotate ground truth for music traditions with similar characteristics to flamenco in order to evaluate the suitability of our approach for a larger variety of genres. 

% or


\section*{Acknowledgment}
This research was funded by the PhD fellowship of the Department of Information and Communication Technologies (DTIC), Universitat Pompeu Fabra and the projects SIGMUS (TIN2012-36650) and COFLA II (P12-TIC-1362).

\ifCLASSOPTIONcaptionsoff
  \newpage
\fi



%\bibliographystyle{IEEEtran}

\newpage

\end{document}